\documentclass[aps,pra,showpacs,twocolumn]{revtex4}
\usepackage{amsfonts}
\usepackage{amssymb}
\usepackage{amsmath}
\usepackage{graphicx}
\usepackage{epsfig}

\begin{document}

\title{Generation of GHZ and W states for stationary qubits in spin network
via resonance scattering}
\author{L. Jin, and Z. Song$^{\dag}$ }
\affiliation{Department of Physics, Nankai University, Tianjin 300071, China}

\begin{abstract}
We propose a simple scheme to establish entanglement among stationary qubits
based on the mechanism of resonance scattering between them and a
single-spin-flip wave packet in designed spin network. It is found that
through the natural dynamical evolution of an incident single-spin-flip wave
packet in a spin network and the subsequent measurement of the output
single-spin-flip wave packet, multipartite entangled states among $n$
stationary qubits, Greenberger-Horne-Zeilinger (GHZ) and W states can be
generated with success probabilities $P_{GHZ}=2/|1+t^{-n}|^{2} $ and $%
P_{W}=\left\vert t\right\vert ^{2}/n$ respectively, where $t$ is the
transmission amplitude of the near-resonance scattering.
\end{abstract}

\maketitle

\section{Introduction}

In quantum information science it is a crucial problem to develop techniques
for generating entanglement among stationary qubits. Entanglement as unique
feature of quantum mechanics\ can be used not only to test fundamental
quantum-mechanical principles \cite{Bell, Greenberger}, but to play a
central role in applications \cite{Ekert, Deutsch, Bennett}. Especially,
multipartite entanglement has been recognized as a powerful resource in
quantum information processing and communication. There are two typical
multipartite entangled states, Greenberger-Horne-Zeilinger (GHZ) and W
states, which are usually referred to as maximal entanglement. Numerous
protocols for the preparation of such states have been proposed \cite{Cirac,
Gerry, Hagley, Cabrillo, Bose, Lange, Rauschenbeutel, Zheng, Feng, Simon,
Zou, Duan, SongJ, Su, SongHS}.

Most of them are scattering-based schemes which utilize two processes: the
natural dynamic process of an always on system and the final project process
carried out by a subsequent measurement. Another feature of such kind of
schemes is that there are two kinds of qubits involved in: target qubits and
flying qubit. The target qubits are the main entities that will be entangled
by the above two processes, which are usually stationary and can be realized
by atoms, impurities, or other quantum devices. The flying qubit is a\
mediator to establish the entanglement among the target qubits via the
interaction between them, which is usually realized by photon or mobile
electron.

In this sense, the type of interaction between stationary qubits and the
flying qubit as a mediator and the transfer of the flying qubit are crucial
for the efficiency of the entanglement creation. In general, such two
processes are mutually exclusive. The scattering between stationary and
flying qubits can convert information between them, while it also reduces
the fidelity of the flying qubit, which will affect the efficiency of the
entanglement, especially for multi-particle system. It is still a challenge
to create entanglement among massive, or stationary qubits.

In this paper, we consider whether it is possible to use an arrangement of
qubits, a spin network, to generate multipartite entanglement among
stationary qubits via scattering process. We introduce a scheme that allow
the generation of the GHZ and W states of stationary qubits in spin
networks. In the proposed scheme, the flying qubit is a Gaussian type
single-flip moving wave packet on the ferromagnetic background, which can
propagate freely in $XY$ chain. The stationary qubit is consisted of two
spins coupled by Ising type interaction with strength $J^{z}$. A single spin
flip can be confined inside such two spins by local magnetic field $h$ to
form a\ double-dot (DD) qubit. The system of an $XY$ spin chain with a DD
qubit embedded in exhibits a novel feature under the resonance scattering
condition $h=J^{z}$, that a single-flip moving wave packet can completely
pass over a DD qubit and switch it from state $\left\vert 0\right\rangle $
to $\left\vert 1\right\rangle $ simultaneously. We show that the scattering
between a flying qubit and a DD qubit can induce the entanglement between
them and the operation on the DD qubit can be performed by the measurement
of the output flying qubit. It allows simple schemes for generation of
multipartite entanglement, such as GHZ and W states by simply-designed spin
networks. We also investigate the influence of near-resonance effects on the
success probabilities of the schemes. It is found that the success
probabilities are $P_{GHZ}=$ $2\left\vert 1+t^{-n}\right\vert ^{2}$ and $%
P_{W}=\left\vert t\right\vert ^{2}/n$\ for the generation of GHZ and W
states, respectively. Here $t$ is the transmission probability amplitude for
a single DD qubit and $n$ is the number of the DD qubits.

This paper is organized as follows. In Sec. II the DD qubit and spin network
are presented. In Sec. III we investigate the resonance-scattering process
between the flying and stationary qubits. Sec. IV and V are devoted to the
application of the resonance scattering on schemes of creating GHZ and W
states. Section VI is the summary and discussion.

\section{Double-dot qubit}

The spin network we consider in this paper is consisted of spins connected
via the $XXZ$ interaction. The Hamiltonian is
\begin{eqnarray}
H &=&-\sum_{\left\langle i,j\right\rangle }\left[ \frac{J_{ij}^{\bot }}{2}%
(\sigma _{i}^{+}\sigma _{j}^{-}+\mathrm{H.c.})+J_{ij}^{z}\sigma
_{i}^{z}\sigma _{j}^{z}\right] \\
&&+\sum_{i}h_{i}\sigma _{i}^{z},  \notag
\end{eqnarray}%
where $\sigma _{i}^{\pm }=\sigma _{i}^{x}\pm i\sigma _{i}^{y}$, and $\sigma
_{i}^{\alpha }$ $(\alpha =x,y,z)$ are the Pauli spin matrices for the spin
at site $i$. The total $z$-component of spin, or the number of spin flips on
the ferromagnetic background, is conserved as it commutes with the
Hamiltonian.\ For $J_{ij}^{z}=0$, it reduces to $XY$ spin network, which has
received a wide study for the purpose of quantum state transfer and creating
entanglement between distant qubits by using the natural dynamics \cite{Bose
Rev}. For $J_{ij}^{z}=J_{ij}^{\bot }$, the Hamiltonian describes isotropic
Heisenberg model. In the antiferromagnetic regime ($J_{ij}^{z}\prec 0$), a
ladder geometry spin network, a gapped system \cite{LiY PRA}, has been
employed as a data bus for the swapping operation and generation of
entanglement between two distant stationary qubits. It has been shown that a
moving wave packet can act as a flying qubit \cite{Osborne, Shi PRA, Yang
PRA} like photon in a fiber. On the other hand, the analogues of optical
device, beam splitter can be fabricated in quantum networks of bosonic \cite%
{Plenio}, spin and ferimonic systems \cite{Yang EPJB}.

In this paper, we consider a new type of qubit, double-dot qubit, which can
be embedded in such spin networks. A DD qubit consists of two ordinary spins
at sites $d$ and $d+1,$ connected via Ising type interaction in the form%
\begin{equation}
H_{d}=-J^{z}\sigma _{d}^{z}\sigma _{d+1}^{z}+h\sum_{i=d,d+1}\sigma _{i}^{z}.
\label{H_dd}
\end{equation}%
When such kinds of two spins are embedded in the spin networks with $%
\left\vert h\right\vert \gg J_{ij}^{\bot }$ and $h_{i}=0$, a spin flip is
confined within it and forms a DD qubit with the notations $\left\vert
0\right\rangle _{d}=\left\vert \downarrow \right\rangle _{d}\left\vert
\uparrow \right\rangle _{d+1}$ and $\left\vert 1\right\rangle =\left\vert
\uparrow \right\rangle _{d}\left\vert \downarrow \right\rangle _{d+1}$. We
will show that such a new type of qubit has a novel feature when it
interacts with another spin flip in the spin networks.

\begin{figure}[tbp]
\includegraphics[ bb=174 356 417 523 , width=7cm , clip]{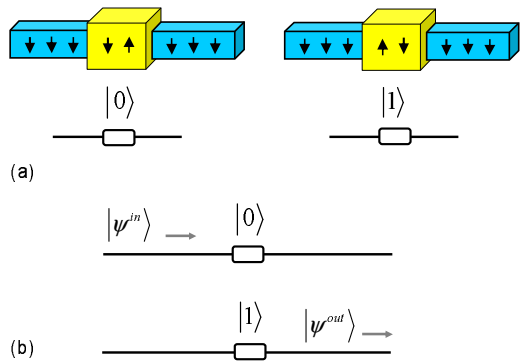}
\caption{\textit{(Color online) The basic building block of the proposed
scheme. (a) Schematic illustration for a DD qubit embedded in a spin chain,
which consists of two spins with only Ising interaction. The local external
magnetic field can confine a spin flip in the two spins to form a DD qubit
with states }$\left\vert 0\right\rangle $\textit{\ and }$\left\vert
1\right\rangle $\textit{. (b) Schematic illustration of the resonance
scattering process between a single-spin-flip wave packet and a DD qubit.
Under the resonance scattering condition, an incident wave packet can
totally pass through a DD qubit and switch it from state }$\left\vert
0\right\rangle $\textit{\ to state }$\left\vert 1\right\rangle .$\textit{\ }}
\label{fig1}
\end{figure}

\section{Resonant scattering}

The main building block in the spin network of our scheme is the DD qubit.
It acts as a massive or stationary qubit, like atoms or ions in
cavity-QED-based schemes. To demonstrate the property of a DD qubit in a
spin chain, we investigate a small system of $4$-site, a DD qubit connecting
to two spins. In order to provide a clear exposition, we firstly assume a
specific coupling configuration with $h=J^{z}$, which leads to the following
$4$-site Hamiltonian%
\begin{eqnarray}
H_{s} &=&-J^{\bot }\sum_{l=1,3}\left( \sigma _{l}^{x}\sigma
_{l+1}^{x}+\sigma _{l}^{y}\sigma _{l+1}^{y}\right)  \label{H_s} \\
&&-h\sigma _{2}^{z}\sigma _{3}^{z}+h\left( \sigma _{2}^{z}+\sigma
_{3}^{z}\right) .  \notag
\end{eqnarray}%
There is a quasi-invariant subspace with the diagonal energy being $h$ and
under the condition $h\gg \left\vert J^{\bot }\right\vert $, which is
spanned by basis%
\begin{eqnarray}
\left\vert \varphi _{1}\right\rangle &=&\left\vert \uparrow \right\rangle
_{1}\left\vert \downarrow \right\rangle _{2}\left\vert \uparrow
\right\rangle _{3}\left\vert \downarrow \right\rangle _{4},  \notag \\
\left\vert \varphi _{2}\right\rangle &=&\left\vert \downarrow \right\rangle
_{1}\left\vert \uparrow \right\rangle _{2}\left\vert \uparrow \right\rangle
_{3}\left\vert \downarrow \right\rangle _{4}, \\
\left\vert \varphi _{3}\right\rangle &=&\left\vert \downarrow \right\rangle
_{1}\left\vert \uparrow \right\rangle _{2}\left\vert \downarrow
\right\rangle _{3}\left\vert \uparrow \right\rangle _{4}.  \notag
\end{eqnarray}%
The matrix of the Hamiltonian in this subspace reads%
\begin{equation}
\left[
\begin{array}{ccc}
h & -\frac{J^{\bot }}{2} &  \\
-\frac{J^{\bot }}{2} & h & -\frac{J^{\bot }}{2} \\
& -\frac{J^{\bot }}{2} & h%
\end{array}%
\right]
\end{equation}%
with eigenstates $\left\vert \psi _{i}\right\rangle $ and eigen energies $%
\varepsilon _{i}$\ $(i=1,2,3)$\ being
\begin{eqnarray}
\left\vert \psi _{1}\right\rangle &=&\frac{1}{\sqrt{2}}\left( \left\vert
\varphi _{1}\right\rangle -\left\vert \varphi _{3}\right\rangle \right)
,\varepsilon _{1}=h; \\
\left\vert \psi _{2,3}\right\rangle &=&\frac{1}{2}\left( \left\vert \varphi
_{1}\right\rangle \mp \sqrt{2}\left\vert \varphi _{2}\right\rangle
+\left\vert \varphi _{3}\right\rangle \right) ,  \notag \\
\varepsilon _{2,3} &=&\pm \frac{1}{\sqrt{2}}J^{\bot }+h.  \notag
\end{eqnarray}%
Obviously, in the invariant subspace, such a $4$-site system acts as a
normal $3$-site system. Note that the DD qubit as the center of the $3$-site
system can be in two different states $\left\vert 0\right\rangle $ or $%
\left\vert 1\right\rangle $ while another spin flip is at left or right
site. Thus a time evolution process can accomplish the transformation $%
\left\vert \uparrow \right\rangle _{1}\left\vert \downarrow \right\rangle
_{2}\left\vert \uparrow \right\rangle _{3}\left\vert \downarrow
\right\rangle _{4}\longrightarrow \left\vert \downarrow \right\rangle
_{1}\left\vert \uparrow \right\rangle _{2}\left\vert \downarrow
\right\rangle _{3}\left\vert \uparrow \right\rangle _{4}$ or $\left\vert
\uparrow \right\rangle _{1}\left\vert 0\right\rangle _{d}\left\vert
\downarrow \right\rangle _{4}\longrightarrow \left\vert \downarrow
\right\rangle _{1}\left\vert 1\right\rangle _{d}\left\vert \uparrow
\right\rangle _{4}$\ with 100\% success probability. Such a feature is
desirable for the quantum information processing. It is because that, on one
hand, a spin flip passing over a DD qubit can operate the qubit state; and
on the other hand, the state of a DD qubit can indicate whether there is a
spin flip passing over it. A similar transformation has been proposed
through the cavity input-output process in adiabatic limit \cite{Lee}. A
particular merit of the present scheme is that it is based on a natural
dynamic process rather than an adiabatic process.

Now we consider the dynamic process of the interaction between a moving wave
packet and a DD qubit. We embed a DD qubit into a chain as illustrated in
Fig. \ref{fig1}(a). It has been shown that a single-spin-flip wave packet in
the form \cite{Osborne, Yang PRA}%
\begin{equation}
\left\vert \phi (\frac{\pi }{2},N_{c})\right\rangle =\frac{1}{\sqrt{\Omega }}%
\sum_{j}e^{-\frac{\alpha ^{2}}{2}(j-N_{c})^{2}+i\frac{\pi }{2}j}\sigma
_{j}^{+}\left\vert Vac\right\rangle  \label{single P}
\end{equation}%
can propagate along a uniform spin chain without spreading approximately,
where the vacuum state is fully ferromagnetic state $\left\vert
Vac\right\rangle =\prod\nolimits_{l=1}\left\vert \downarrow \right\rangle
_{l}$. Here $\Omega =\sum_{1}^{N}\exp (-\alpha ^{2}(j-N_{c})^{2}/2)$ is the
normalization factor, $N_{c}$\ is the center of the wave packet at $t=0$ and
$N$ is the number of sites of the chain. At time $t$, it will evolves to $%
\left\vert \phi (\pi /2,N_{c}+J^{\bot }t)\right\rangle $.\ Let us firstly
assume that initially the qubit is in the state $\left\vert 0\right\rangle
_{d}$, while a wave packet of type (\ref{single P}) $\left\vert \phi (\pi
/2,N_{c}\prec d)\right\rangle \equiv $ $\left\vert \phi (\pi
/2,L)\right\rangle $ is coming from the left. Similarly, we define $%
\left\vert \phi (\pm \pi /2,N_{c}\succ d+2)\right\rangle $ $\equiv
\left\vert \phi (\pm \pi /2,R)\right\rangle $, $\left\vert \phi (-\pi
/2,N_{c}\prec d)\right\rangle $ $\equiv \left\vert \phi (-\pi
/2,L)\right\rangle $ to denote a transmitted or reflected wave packet after
scattering. In the strong local field regime, $h\gg \left\vert J^{\bot
}\right\vert $, the spin flip is firmly confined in the DD qubit. From the
analysis of the above $4$-site system, which is called the resonant case
with $h=J^{z}$, the wave packet will pass freely through the DD qubit.
Comparing to the case without the embedded DD qubit, the output wave packet
gets a forward shift with a lattice space, while switches the DD qubit from $%
\left\vert 0\right\rangle _{d}$\ to $\left\vert 1\right\rangle _{d}$, i.e.,

\begin{equation}
\left\vert \phi (\frac{\pi }{2},L)\right\rangle \left\vert 0\right\rangle
_{d}\longrightarrow \left\vert \phi (\frac{\pi }{2},R)\right\rangle
\left\vert 1\right\rangle _{d}.  \label{tran1}
\end{equation}%
In contrast, if the qubit is in state $\left\vert 1\right\rangle $, the
scattering process is

\begin{equation}
\left\vert \phi (\frac{\pi }{2},L)\right\rangle \left\vert 1\right\rangle
_{d}\longrightarrow \left\vert \phi (-\frac{\pi }{2},L)\right\rangle
\left\vert 1\right\rangle _{d},  \label{tran2}
\end{equation}%
i.e., the incoming wave packet is totally reflected and maintains the qubit
to be in state $\left\vert 1\right\rangle _{d}$. Interestingly, the states
of wave packet and the DD qubit\ are both\ altered through this process,
i.e., being shifted with a lattice space. However, such a shift brings about
totally different effects on the DD qubit and the wave packet, respectively:
it switches the DD qubit from $\left\vert 0\right\rangle _{d}$\ to $%
\left\vert 1\right\rangle _{d}$, but does not\ alter the wave packet in the
same manner\ as \textit{classical} perfect elastic collision.\ It is worthy
to point out that the DD qubit and the wave packet are not\ entangled in
such resonant case. In the case of non-resonance $h\neq J^{z}$, or initially
the DD qubit and/or the incident wave packet are in a superposition states as

\begin{eqnarray}
\left\vert \nearrow \right\rangle _{d} &=&\alpha \left\vert 0\right\rangle
_{d}+\beta \left\vert 1\right\rangle _{d}, \\
\left\vert \nearrow \right\rangle _{in} &=&\alpha \left\vert \phi (\frac{\pi
}{2},L)\right\rangle +\beta \left\vert Vac\right\rangle ,  \notag
\end{eqnarray}%
where $\alpha $ and $\beta $ are arbitrary coefficients satisfying $%
\left\vert \alpha \right\vert ^{2}+\left\vert \beta \right\vert ^{2}=1$.

\begin{figure}[tbp]
\includegraphics[ bb=15 584 222 744, width=8.0 cm, clip]{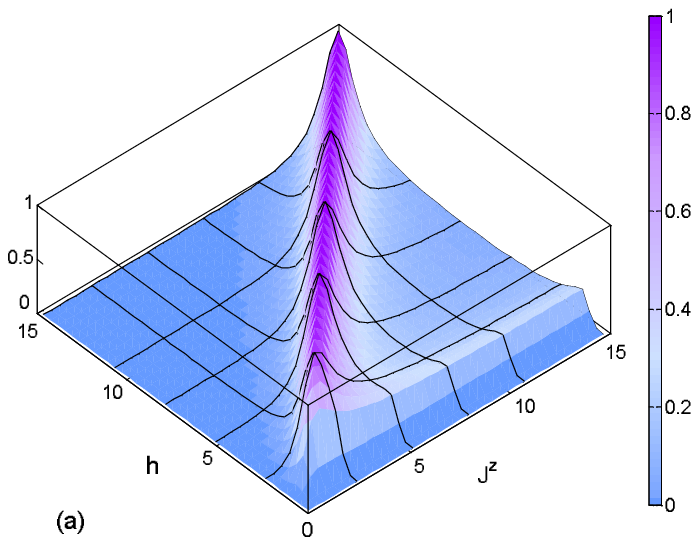} %
\includegraphics[ bb=12 511 295 714, width=7.0 cm, clip]{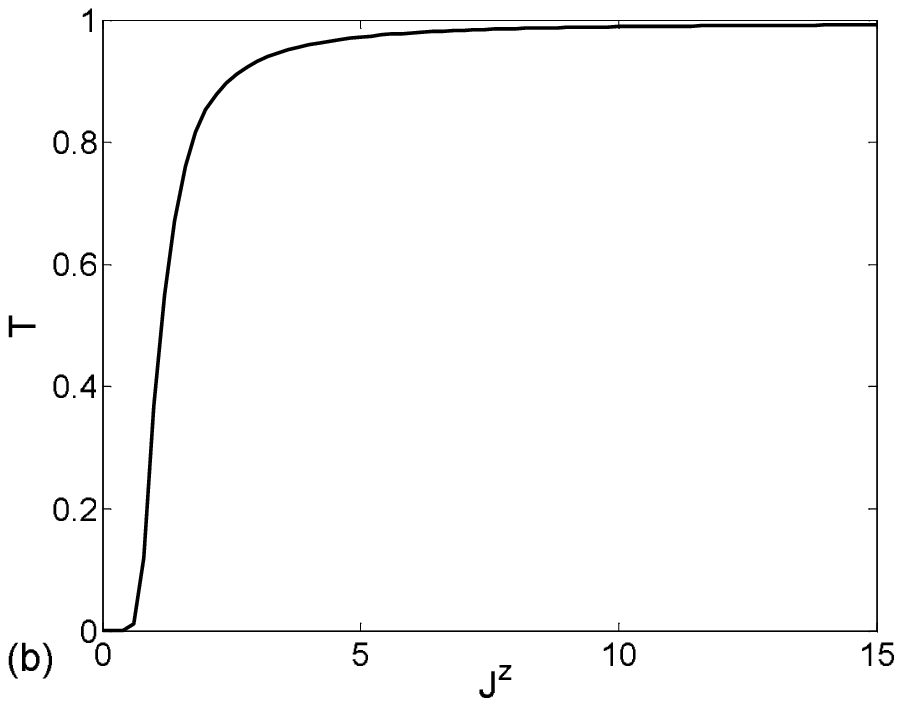}
\caption{\textit{(Color online) Profiles of transmission coefficient }$T$%
\textit{\ as a function of }$h$\textit{\ and }$J^{z}$\textit{\ in the unit
of }$J^{\bot }$\textit{, obtained by numerical simulations. (a) 3-D plot of
the function }$T(h,J^{z})$\textit{. It indicates that the transmission
coefficient gets the maxima under the resonance scattering condition }$%
h=J^{z}$\textit{. (b) The plot of the transmission coefficient under the
resonance scattering condition. It shows that the transmission coefficient
is close to }$1$\textit{\ in the region }$h\sim J^{z}\succ \left\vert
5J^{\bot }\right\vert $\textit{.}}
\label{fig2}
\end{figure}

In practice, the difference between $h$ and $J^{z}$ will leads to the
reflection of the incident wave packet. Then the scattering process can be
expressed as

\begin{equation}
\left\vert \phi (\frac{\pi }{2},L)\right\rangle \left\vert 0\right\rangle
_{d}\longrightarrow r\left\vert \phi (-\frac{\pi }{2},L)\right\rangle
\left\vert 0\right\rangle _{d}+t\left\vert \phi (\frac{\pi }{2}%
,R)\right\rangle \left\vert 1\right\rangle _{d}
\end{equation}%
with $\left\vert r\right\vert ^{2}+\left\vert t\right\vert ^{2}=1$. The
transmission coefficient $T(J^{z},h)=\left\vert t\right\vert ^{2}$ is a
crucial factor in the following schemes for quantum information processing.
On the other hand, the strength of the local field $h$ also results the
spreading of the spin flip from state $\left\vert 1\right\rangle _{d}$\ and
reduces $T(J^{z},h)$.\ We perform numerical simulation for the scattering
process in order to investigate such phenomenon.\ Numerical result for $%
T(J^{z},h)$\ with $\alpha =4/15$ is plotted in Fig. \ref{fig2}. It shows
that the transmission coefficient is close to $1$ if $h\sim J^{z}\succ
5\left\vert J^{\bot }\right\vert $, which is feasible in practice. It
ensures that a spin network with an embedded DD qubit in a spin chain can
perform the transformation (\ref{tran1}, \ref{tran2}) via a natural dynamic
process rather than an adiabatic process.

\section{Generation of GHZ state}

Now we focus on the practical application of resonant scattering effect on
the quantum information processing. As mentioned above, although the totally
transmitted wave packet switches the state of the DD qubit from $\left\vert
0\right\rangle _{d}$\ to $\left\vert 1\right\rangle _{d}$, there is no
entanglement between the DD and the wave packet arising from such a process.
However, if the incoming wave packet is not polarized in $z$ direction, but
in the form

\begin{equation}
\left\vert \nearrow \right\rangle _{in}=\alpha \left\vert \phi (\frac{\pi }{2%
},L)\right\rangle +\beta \left\vert Vac\right\rangle ,  \label{WP_in}
\end{equation}%
where $\alpha $ and $\beta $ are restricted to be real for simplicity in the
following context, the entanglement between the DD qubit and the scattered
wave packet can be established. Actually, the corresponding resonant
scattering process can be expressed as

\begin{equation}
\left\vert \nearrow \right\rangle _{in}\left\vert 0\right\rangle
_{d}\longrightarrow \alpha \left\vert 1\right\rangle _{d}\left\vert \phi (%
\frac{\pi }{2},R)\right\rangle +\beta \left\vert 0\right\rangle
_{d}\left\vert Vac\right\rangle .  \label{initial_1}
\end{equation}%
The reduced density matrix of the final state in the basis $\{\left\vert
1\right\rangle _{d}\left\vert Vac\right\rangle $, $\left\vert 1\right\rangle
_{d}\left\vert \phi (\pi /2,R)\right\rangle $, $\left\vert 0\right\rangle
_{d}\left\vert Vac\right\rangle $, $\left\vert 0\right\rangle _{d}\left\vert
\phi (\pi /2,R)\right\rangle \}$ is%
\begin{equation}
\left[
\begin{array}{cccc}
0 &  &  &  \\
& \alpha ^{2} & \alpha \beta &  \\
& \alpha \beta & \beta ^{2} &  \\
&  &  & 0%
\end{array}%
\right] ,
\end{equation}%
which has concurrence $C=2\left\vert \alpha \beta \right\vert $. For $\alpha
=\beta =1/\sqrt{2}$, the concurrence between them reaches to the maximum $%
C_{\max }=1$. A flying qubit (\ref{WP_in}) can be generated from wave packet
(\ref{single P}) via a $1\times 2$-beam or $Y$-beam splitter \cite{Yang EPJB}%
. We consider the spin network with the geometry of two connected $1\times 2$%
-beam splitters. These two $Y$-beam splitters are characterized by $\left(
\alpha \text{, }\beta \right) $ and $\left( \alpha ^{\prime }\text{, }\beta
^{\prime }\right) $ respectively, which are schematically shown in Fig. \ref%
{fig3}(a). There is a DD qubit embedded in one of the two arms. Now we
consider the dynamic process with the initial state being $\left\vert \psi
^{in}\right\rangle \left\vert 0\right\rangle _{d}$, where $\left\vert \psi
^{in}\right\rangle =\left\vert \phi (\pi /2,-\infty )\right\rangle $,\
denotes an incoming wave packet along the left chain. In the first step,
through the beam splitter $\left( \alpha \text{, }\beta \right) $,\ $%
\left\vert \psi ^{in}\right\rangle $\ is divided into two wave packets $%
\left\vert \phi (\pi /2,L)\right\rangle _{A}$\ and $\left\vert \phi (\pi
/2)\right\rangle _{B}$\ along $A$ and $B$ arms, respectively. In the second
step, sub-wave packet $\left\vert \phi (\pi /2,L)\right\rangle _{A}$ passes
over DD qubit and becomes $\left\vert \phi (\pi /2,R)\right\rangle _{A}$,
while sub-wave packet $\left\vert \phi (\pi /2)\right\rangle _{B}$
propagates along $B$ arm and meets $\left\vert \phi (\pi /2,R)\right\rangle
_{A}$ at the joint of beam splitter $\left( \alpha ^{\prime }\text{, }\beta
^{\prime }\right) $. In the third step, wave packets $\left\vert \phi (\pi
/2,R)\right\rangle _{A}$\ and $\left\vert \phi (\pi /2)\right\rangle _{B}$\
are reflected and divided by beam splitter $\left( \alpha ^{\prime }\text{, }%
\beta ^{\prime }\right) $, and contribute to the output wave packet $%
\left\vert \psi ^{out}\right\rangle =\left\vert \phi (\pi /2,\infty
)\right\rangle $. Then the whole process can be expressed as

\begin{eqnarray}
&&\left\vert \psi ^{in}\right\rangle \left\vert 0\right\rangle _{d} \\
&&\overset{\text{step 1}}{\longrightarrow }\alpha \left\vert \phi (\frac{\pi
}{2},L)\right\rangle _{A}\left\vert 0\right\rangle _{d}+\beta \left\vert
\phi (\frac{\pi }{2})\right\rangle _{B}\left\vert 0\right\rangle _{d}  \notag
\\
&&\overset{\text{step 2}}{\longrightarrow }\alpha \left\vert 1\right\rangle
_{d}\left\vert \phi (\frac{\pi }{2},R)\right\rangle _{A}+\beta \left\vert
0\right\rangle _{d}\left\vert \phi (\frac{\pi }{2})\right\rangle _{B}  \notag
\\
&&\overset{\text{step 3}}{\longrightarrow }\left( \alpha ^{\prime }\alpha
\left\vert 1\right\rangle _{d}+\beta ^{\prime }\beta \left\vert
0\right\rangle _{d}\right) \left\vert \psi ^{out}\right\rangle  \notag \\
&&+\left( \beta ^{\prime 2}\alpha \left\vert 1\right\rangle _{d}-\alpha
^{\prime }\beta ^{\prime }\beta \left\vert 0\right\rangle _{d}\right)
\left\vert \phi (-\frac{\pi }{2},R)\right\rangle _{A}  \notag \\
&&+\left( \alpha ^{\prime 2}\beta \left\vert 0\right\rangle _{d}-\alpha
^{\prime }\beta ^{\prime }\alpha \left\vert 1\right\rangle _{d}\right)
\left\vert \phi (-\frac{\pi }{2})\right\rangle _{B}.  \notag
\end{eqnarray}%
From step 2 to step 3 we have used formulas (\ref{Alfa in}, \ref{Beta in})
derived in Appendix A. When $\left\vert \psi ^{out}\right\rangle $ is
measured in the output lead, the operation
\begin{equation}
\left\vert 0\right\rangle _{d}\longrightarrow \alpha ^{\prime }\alpha
\left\vert 1\right\rangle _{d}+\beta ^{\prime }\beta \left\vert
0\right\rangle _{d}
\end{equation}%
is implemented. The success probability of this operation is $\left( \alpha
\alpha ^{\prime }\right) ^{2}+\left( \beta \beta ^{\prime }\right) ^{2}$. In
the optimal case with $\alpha =\beta =\alpha ^{\prime }$ $=\beta ^{\prime
}=1/\sqrt{2}$, we can perform the operation $\left\vert 0\right\rangle
_{d}\longrightarrow \left( \left\vert 1\right\rangle _{d}+\left\vert
0\right\rangle _{d}\right) /\sqrt{2}$ by the process of resonant scattering
and subsequent measurement with the success probability up to $0.5$. The
measurement of $\left\vert \psi ^{out}\right\rangle $\ can be implemented by
embedding another DD qubit to record the passing of the output wave packet.

Now we consider the case of multiple DD qubits embedded in $A$ arm, which is
schematically illustrated in Fig. \ref{fig3}(b). All the $n$ stationary DD
qubits are prepared initially in state $\left\vert 0\right\rangle ^{\otimes
n}=\prod\nolimits_{l=1}^{n}\left\vert 0\right\rangle _{l}$.\ Similarly, we
have
\begin{eqnarray}
&&\left\vert \psi ^{in}\right\rangle \left\vert 0\right\rangle ^{\otimes n}
\\
&&\overset{\text{step 1}}{\longrightarrow }\alpha \left\vert \phi (\pi
/2,L)\right\rangle _{A}\left\vert 0\right\rangle ^{\otimes n}+\beta
\left\vert \phi (\frac{\pi }{2})\right\rangle _{B}\left\vert 0\right\rangle
^{\otimes n}  \notag \\
&&\overset{\text{step 2}}{\longrightarrow }\alpha \left\vert \phi (\pi
/2,R)\right\rangle _{A}\left\vert 1\right\rangle ^{\otimes n}+\beta
\left\vert \phi (\frac{\pi }{2})\right\rangle _{B}\left\vert 0\right\rangle
^{\otimes n}  \notag \\
&&\overset{\text{step 3}}{\longrightarrow }\left( \alpha ^{\prime }\alpha
\left\vert 1\right\rangle ^{\otimes n}+\beta ^{\prime }\beta \left\vert
0\right\rangle ^{\otimes n}\right) \left\vert \psi ^{out}\right\rangle
\notag \\
&&+\left( \beta ^{\prime 2}\alpha \left\vert 1\right\rangle ^{\otimes
n}-\alpha ^{\prime }\beta ^{\prime }\beta \left\vert 0\right\rangle
^{\otimes n}\right) \left\vert \phi (-\frac{\pi }{2},R)\right\rangle _{A}
\notag \\
&&+\left( \alpha ^{\prime 2}\beta \left\vert 0\right\rangle ^{\otimes
n}-\alpha ^{\prime }\beta ^{\prime }\alpha \left\vert 1\right\rangle
^{\otimes n}\right) \left\vert \phi (-\frac{\pi }{2})\right\rangle _{B},
\notag
\end{eqnarray}%
with the notation $\left\vert 1\right\rangle ^{\otimes
n}=\prod\nolimits_{l=1}^{n}\left\vert 1\right\rangle _{l}$.

In the optimal case with $\alpha =\beta =\alpha ^{\prime }$ $=\beta ^{\prime
}=1/\sqrt{2}$, we can perform the operation
\begin{equation}
\left\vert 0\right\rangle ^{\otimes n}\longrightarrow \frac{1}{\sqrt{2}}%
\left( \left\vert 1\right\rangle ^{\otimes n}+\left\vert 0\right\rangle
^{\otimes n}\right)
\end{equation}%
by the subsequent measurement with the success probability up to $0.5$. Then
by using natural dynamics and subsequent measurement, multipartite entangled
GHZ state can be generated. This provides a simple way of entangling $n$
stationary qubits through scattering with a flying qubit. In the
near-resonance scattering case, the transmission probability amplitude $t$
will effect the success probability to be
\begin{equation}
P_{GHZ}=\frac{2}{\left\vert 1+t^{-n}\right\vert ^{2}},  \label{P_GHZ}
\end{equation}%
under the optimal conditions $\alpha =\alpha ^{\prime }=\sqrt{1/\left(
1+t^{n}\right) }$, $\beta =\beta ^{\prime }=\sqrt{t^{n}/\left(
1+t^{n}\right) }$. Note that the success probability is reduced
exponentially as the number of qubits increases.

\section{Generation of W state}

\begin{figure}[tbp]
\includegraphics[ bb=199 296 407 542, width=7.0 cm, clip]{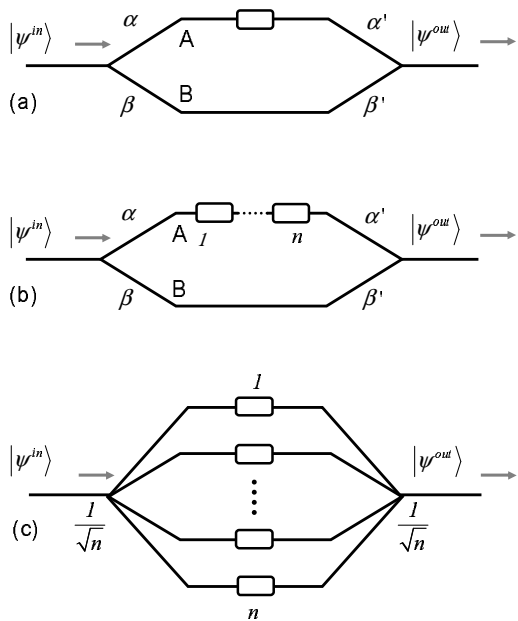}
\caption{\textit{Schematic illustrations for the scheme of entanglement
generation in spin network. (a) Realization of operation }$\left\vert
0\right\rangle \longrightarrow $\textit{\ }$(\left\vert 1\right\rangle
+\left\vert 0\right\rangle /\protect\sqrt{2}$\textit{. (b) Generation of the
GHZ state. (c) Generation of the W state.}}
\label{fig3}
\end{figure}

Now we turn to the scheme of the generation of another type of multipartite
entangled state, W state. The configuration of the spin network we utilized
consists of two $1\times n$-beam splitters\ with one DD qubit embedded in
each parallel arm in the same way, as illustrated in Fig. \ref{fig3}(c).

We start our analysis by considering $n=2$ case with two $1\times 2$-beam
splitters being characterized by $\left( \alpha \text{, }\beta \right) $ and
$\left( \alpha ^{\prime }\text{, }\beta ^{\prime }\right) $ respectively.
Denoting the qubit states of two DD qubits embedded in arms $A$ and $B$ as $%
\left\vert 0,1\right\rangle _{A}$ and $\left\vert 0,1\right\rangle _{B}$
respectively, the dynamic process can be written as

\begin{eqnarray}
&&\left\vert \psi ^{in}\right\rangle \left\vert 0\right\rangle
_{A}\left\vert 0\right\rangle _{B} \\
&&\overset{\text{step 1}}{\longrightarrow }\alpha \left\vert \phi (\frac{\pi
}{2},L)\right\rangle _{A}\left\vert 0\right\rangle _{A}\left\vert
0\right\rangle _{B}+\beta \left\vert \phi (\frac{\pi }{2},L)\right\rangle
_{B}\left\vert 0\right\rangle _{A}\left\vert 0\right\rangle _{B}  \notag \\
&&\overset{\text{step 2}}{\longrightarrow }\alpha \left\vert 1\right\rangle
_{A}\left\vert \phi (\frac{\pi }{2},R)\right\rangle _{A}\left\vert
0\right\rangle _{B}+\beta \left\vert 1\right\rangle _{B}\left\vert \phi (%
\frac{\pi }{2},R)\right\rangle _{B}\left\vert 0\right\rangle _{A}  \notag \\
&&\overset{\text{step 3}}{\longrightarrow }\left( \alpha \alpha ^{\prime
}\left\vert 1\right\rangle _{A}\left\vert 0\right\rangle _{B}+\beta \beta
^{\prime }\left\vert 1\right\rangle _{B}\left\vert 0\right\rangle
_{A}\right) \left\vert \psi ^{out}\right\rangle  \notag \\
&&+\left( \alpha \beta ^{\prime 2}\left\vert 1\right\rangle _{A}\left\vert
0\right\rangle _{B}-\beta \alpha ^{\prime }\beta ^{\prime }\left\vert
1\right\rangle _{B}\left\vert 0\right\rangle _{A}\right) \left\vert \phi (-%
\frac{\pi }{2},R)\right\rangle _{A}  \notag \\
&&+\left( \beta \alpha ^{\prime 2}\left\vert 1\right\rangle _{B}\left\vert
0\right\rangle _{A}-\alpha \alpha ^{\prime }\beta ^{\prime }\left\vert
1\right\rangle _{A}\left\vert 0\right\rangle _{B}\right) \left\vert \phi (-%
\frac{\pi }{2},R)\right\rangle _{B}.  \notag
\end{eqnarray}%
In the optimal case with $\alpha =\beta =\alpha ^{\prime }=\beta ^{\prime
}=1/\sqrt{2}$, we can perform the operation
\begin{equation}
\left\vert 0\right\rangle _{A}\left\vert 0\right\rangle _{B}\longrightarrow
\frac{1}{\sqrt{2}}\left( \left\vert 1\right\rangle _{A}\left\vert
0\right\rangle _{B}+\left\vert 0\right\rangle _{A}\left\vert 1\right\rangle
_{B}\right)
\end{equation}%
by the subsequent measurement of the output spin flip with the success
probability up to $0.5$.

Now we extend the above conclusion to $n$-DD qubits case. For a $1\times n$%
-beam splitter, we only consider the simplest case with identical hopping
constant being $J^{\bot }/2\sqrt{n}$ between the input lead and each arm,
which is schematically illustrated in Fig. \ref{fig3}(c). An incident wave
packet will experience the following process. In the first step, $\left\vert
\psi ^{in}\right\rangle $\ is divided into $n$ wave packets through the beam
splitter, with $\left\vert \phi (\pi /2,L)\right\rangle _{l}$\ $\left(
l=1,2,...,n\right) $ being the wave packet along the $l$th arm. In the
second step, every sub-wave packet $\left\vert \phi (\pi /2,L)\right\rangle
_{l}$ passes over the corresponding DD qubit embedded in the $l$th arm, and
switches its state from $\left\vert 0\right\rangle _{l}$ to $\left\vert
1\right\rangle _{l}$ simultaneously. In the third step, all the sub-wave
packets $\left\vert \phi (\pi /2,R)\right\rangle _{l}$\ are reflected and
divided at the node of the right beam splitter, and contribute to the output
wave packet $\left\vert \psi ^{out}\right\rangle =\left\vert \phi (\pi
/2,\infty )\right\rangle $. The whole process can be expressed as
\begin{eqnarray}
&&\left\vert \psi ^{in}\right\rangle \left\vert 0\right\rangle ^{\otimes n}
\\
&&\overset{\text{step 1}}{\longrightarrow }\frac{1}{\sqrt{n}}%
\sum_{l=1}^{n}\left\vert \phi (\frac{\pi }{2},L)\right\rangle
_{l}\prod\nolimits_{l=1}^{n}\left\vert 0\right\rangle _{l}  \notag \\
&&\overset{\text{step 2}}{\longrightarrow }\frac{1}{\sqrt{n}}%
\sum_{l=1}^{n}\left\vert \phi (\frac{\pi }{2},R)\right\rangle _{l}\left\vert
1\right\rangle _{l}\prod\nolimits_{i\neq l}^{n}\left\vert 0\right\rangle _{i}
\notag \\
&&\overset{\text{step 3}}{\longrightarrow }\frac{1}{n}\sum_{l=1}^{n}\left%
\vert 1\right\rangle _{l}\prod\nolimits_{i\neq l}^{n}\left\vert
0\right\rangle _{i}\left\vert \psi ^{out}\right\rangle  \notag \\
&&-\frac{1}{n\sqrt{n}}\sum_{l=1}^{n}\sum_{j=1}^{n}\left\vert \psi
_{j}^{out}\right\rangle \left\vert 1\right\rangle _{l}\prod\nolimits_{i\neq
l}^{n}\left\vert 0\right\rangle _{i}  \notag \\
&&+\sum_{l=1}^{n}\left\vert \psi _{l}^{out}\right\rangle \left\vert
1\right\rangle _{l}\prod\nolimits_{i\neq l}^{n}\left\vert 0\right\rangle _{i}
\notag
\end{eqnarray}%
From step 2 to step 3 we have used the formula (\ref{l_in}) derived in
Appendix B. When $\left\vert \psi ^{out}\right\rangle $ is measured in the
output lead, the operation
\begin{equation}
\left\vert 0\right\rangle ^{\otimes n}\longrightarrow \frac{1}{\sqrt{n}}%
\sum_{l=1}^{n}\left\vert 1\right\rangle _{l}\prod\nolimits_{i\neq
l}^{n}\left\vert 0\right\rangle _{i}
\end{equation}%
is implemented with the success probability $1/n$. Then by using natural
dynamics and subsequent measurement, multipartite entangled W state can be
generated. In the near-resonance scattering case, the transmission
probability amplitude $t$ reduces the success probability to
\begin{equation}
P_{W}=\frac{\left\vert t\right\vert ^{2}}{n}.  \label{P_W}
\end{equation}%

\begin{figure}[tbp]
\includegraphics[ bb=33 604 274 796, width=7.0 cm, clip]{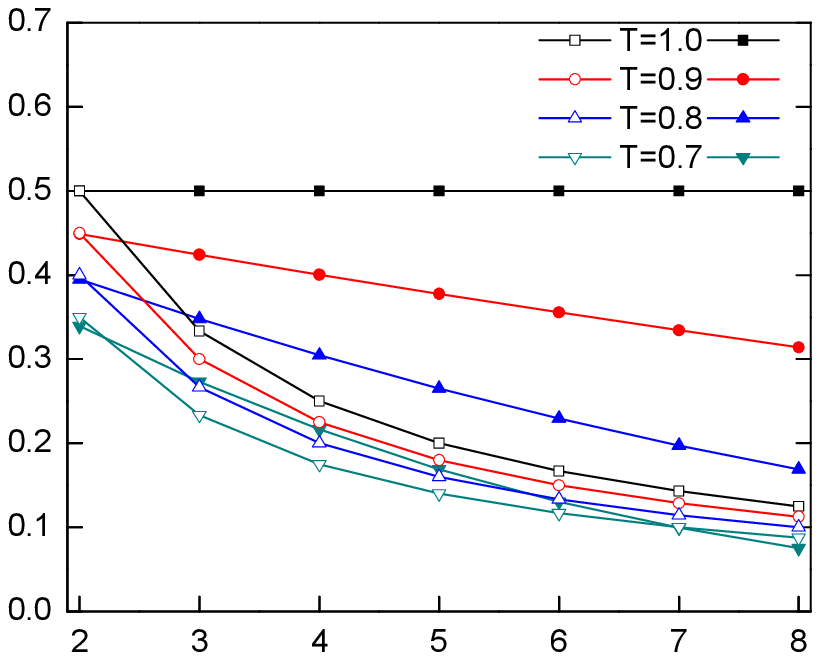}
\caption{\textit{(Color online) Success probabilities for the generation of
GHZ and W states of }$n$\textit{\ qubits in the system with transimission
coefficient }$T$\textit{. }}
\label{GHZ_W}
\end{figure}

As the comparison of the success probabilities of creating GHZ and W states
of $n$ qubits with the transmission coefficient $T$, we plot Eqs. (\ref%
{P_GHZ}) and (\ref{P_W}) in Fig. \ref{GHZ_W} for the cases with $T=1.0$, $%
0.9 $, $0.8$, and $0.7$; $n=2$, $3$, $...$, $8$. It shows that when $T$ is
close to $1$, the difference between $P_{GHZ}$ and $P_{W}$ becomes large as $%
n$ increases.\ As $T$ decreases from $1$, the difference between $P_{GHZ}$
and $P_{W}$ becomes smaller for fixed $n$.

\section{Summary}

We have shown how a spin network can be used to generate multipartite
entanglement among stationary qubits. The key of this scheme is the
alternative of massive or stationary qubit in a spin network, DD qubit. The
resonance scattering between a DD qubit and a SFWP, which acts as the
alternative of a flying qubit in a spin network, allows a perfect
transformation: an incident wave packet can totally pass through a DD qubit
and switch it from state $\left\vert 0\right\rangle $ to state $\left\vert
1\right\rangle $. The resonance scattering condition is investigated
analytically and numerically. It shows that the resonance scattering is
feasible in practice. This ensures that through the natural dynamical
evolution of an incident single-spin-flip wave packet in a spin network and
the subsequent measurement of the output single-spin-flip wave packet,
multipartite entangled states among $n$ stationary qubits, GHZ and W states
can be generated. There are two merits in our scheme. Firstly, the massive
or stationary qubit, DD qubit, is constructed by the element, two neighbor
spins of the spin network, which is applicable to all types of the scalable
multi-qubit systems. Secondly, it is based on a natural dynamic process
rather than an adiabatic process. There is certainly significant potential
for spin networks to find applications in solid state quantum processing and
communication.

We acknowledge the support of the CNSF (grant No. 10874091, 2006CB921205).

\section{Appendix Dynamics of wave packets in beam splitters}

In this appendix, we present the exact results for the dynamics of wave
packets in beam splitters.

\subsection{Appendix A: $1\times 2$ beam splitter}

\begin{figure}[tbp]
\includegraphics[ bb=44 170 416 769, width=3.0 cm, clip]{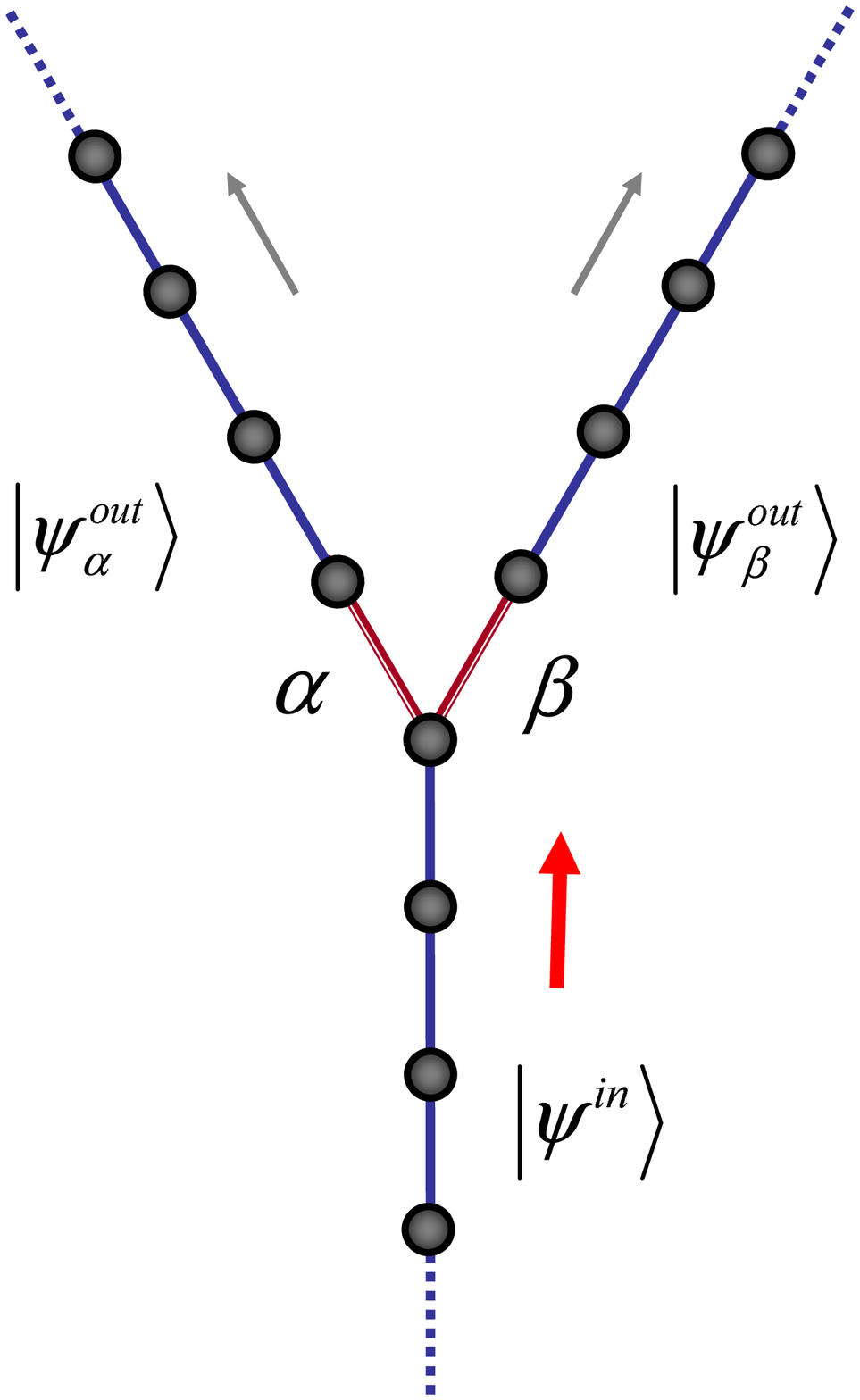} %
\includegraphics[ bb=36 170 409 761, width=3.0 cm, clip]{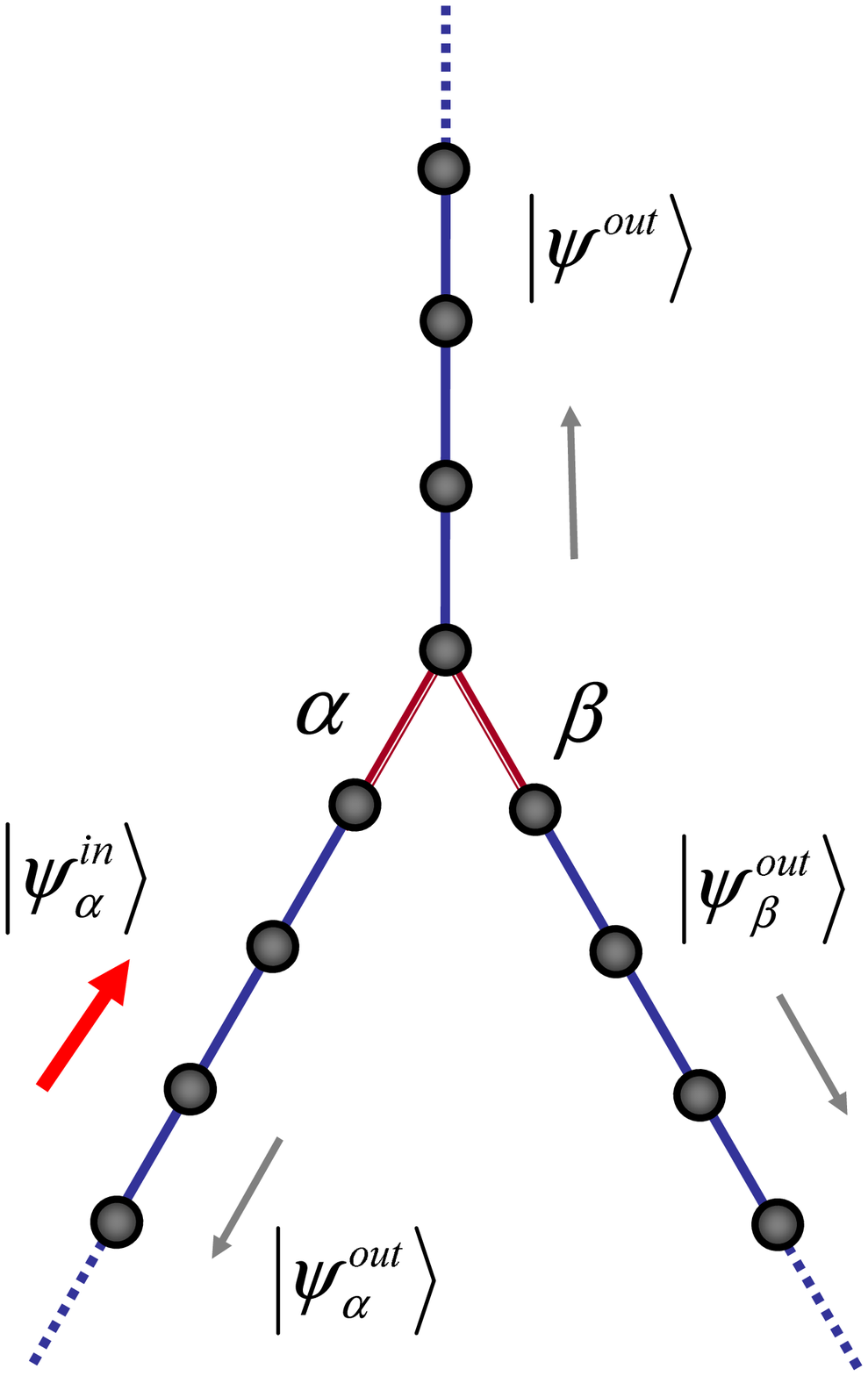}
\caption{\textit{(Color online) Schematic illustration for a }$1\times 2$%
\textit{\ beam splitter. Left panel: an incident wave packet }$\vert \protect%
\psi^{in}\rangle $\textit{\ is split into two sub-wave packets }$\vert
\protect\psi _{\protect\alpha }^{out}\rangle $\textit{\ and }$\vert \protect%
\psi _{\protect\beta }^{out}\rangle $\textit{\ without any reflection. Right
panel: an incident wave packet }$\vert \protect\psi _{\protect\alpha %
}^{in}\rangle $\textit{, along one of the two arms is divided into three
sub-wave packets }$\vert \protect\psi _{\protect\alpha }^{out}\rangle $%
\textit{, }$\vert \protect\psi _{\protect\beta }^{out}\rangle $\textit{\ and
}$\vert \protect\psi ^{out}\rangle $\textit{.}}
\label{fig4}
\end{figure}

In the work of Ref. {\cite{Yang EPJB}} the dynamics of a wave packet in the
spin networks based on the $XY$ model was studied. A $1\times 2$-beam
splitter is consisted of three uniform spin chains with coupling constant $%
J^{\bot }/2$. The connections between three uniform chains are $\alpha
J^{\bot }/2$ and $\beta J^{\bot }/2$\ as in Fig. \ref{fig4}. It has been
shown that under the condition $\alpha ^{2}+\beta ^{2}=1$,\ an input moving
wave packet $\left\vert \psi ^{in}\right\rangle $\ will be divided into two
wave packets $\left\vert \psi _{\alpha }^{out}\right\rangle $ and $%
\left\vert \psi _{\beta }^{out}\right\rangle $\ without any reflection,
which can be expressed as%
\begin{equation}
\left\vert \psi ^{in}\right\rangle \longrightarrow \alpha \left\vert \psi
_{\alpha }^{out}\right\rangle +\beta \left\vert \psi _{\beta
}^{out}\right\rangle .
\end{equation}%
Similarly, the inverse process also holds, i.e.,%
\begin{equation}
\alpha \left\vert \psi _{\alpha }^{in}\right\rangle +\beta \left\vert \psi
_{\beta }^{in}\right\rangle \longrightarrow \left\vert \psi
^{out}\right\rangle ,  \label{A(1)}
\end{equation}%
where states $\left\vert \psi ^{in}\right\rangle $, $\left\vert \psi
_{\alpha }^{in}\right\rangle $, and $\left\vert \psi _{\beta
}^{in}\right\rangle $ ($\left\vert \psi ^{out}\right\rangle $, $\left\vert
\psi _{\alpha }^{out}\right\rangle $, and $\left\vert \psi _{\beta
}^{out}\right\rangle $) represent the wave packets coming in (out) of the
node along the three chains respectively. Contrarily, for two wave packets
along the two arms, which interference destructively at the node, we have
\begin{equation}
\beta \left\vert \psi _{\alpha }^{in}\right\rangle -\alpha \left\vert \psi
_{\beta }^{in}\right\rangle \longrightarrow \beta \left\vert \psi _{\alpha
}^{out}\right\rangle -\alpha \left\vert \psi _{\beta }^{out}\right\rangle .
\label{A(2)}
\end{equation}%
Combining the above Eqs. (\ref{A(1)} and \ref{A(2)}), we have%
\begin{equation}
\left\vert \psi _{\alpha }^{in}\right\rangle \longrightarrow \alpha
\left\vert \psi ^{out}\right\rangle +\beta ^{2}\left\vert \psi _{\alpha
}^{out}\right\rangle -\alpha \beta \left\vert \psi _{\beta
}^{out}\right\rangle ,  \label{Alfa in}
\end{equation}%
and%
\begin{equation}
\left\vert \psi _{\beta }^{in}\right\rangle \longrightarrow \beta \left\vert
\psi ^{out}\right\rangle -\alpha \beta \left\vert \psi _{\alpha
}^{out}\right\rangle +\alpha ^{2}\left\vert \psi _{\beta
}^{out}\right\rangle .  \label{Beta in}
\end{equation}%
Then for a given incident wave packet along any branch of a $1\times 2$ beam
splitter, the probability amplitudes of all the output wave packets can be
obtained exactly.

\subsection{Appendix B: $1\times n$ beam splitter}

Now we consider a $1\times n$ beam splitter consists of one chain of length $%
l_{a}$\ and $n$ arms of length $l_{b}$\ with identical connecting coupling
strength $J^{\bot }/2\sqrt{n}$\ for each arm, which is shown in Fig. \ref%
{fig5}. The Hamiltonian of such quantum network reads

\begin{eqnarray}
\mathcal{H} &=&-\frac{J^{\bot }}{2}\sum_{i=1}^{l_{a}-1}a_{i}^{\dag }a_{i+1}-%
\frac{J^{\bot }}{2\sqrt{n}}\sum_{l=1}^{n}a_{l_{a}}^{\dag }b_{l,1}
\label{Quantum network} \\
&&-\frac{J^{\bot }}{2}\sum_{l=1}^{n}\sum_{j=1}^{l_{b}-1}b_{l,j}^{\dag
}b_{l,j+1}+\mathrm{H.c.},  \notag
\end{eqnarray}%
where $a_{i}^{\dag }$\ and $b_{l,j}^{\dag }$\ are particle operators at site
$i$ of chain $l_{a}$\ and site $j$ of the arm $l$. They can be boson or
fermion operators. The conclusion for such model is available for the
dynamics of a single flip in the analogue of spin network.

\begin{figure}[tbp]
\includegraphics[ bb=41 145 541 741, width=4.0 cm, clip]{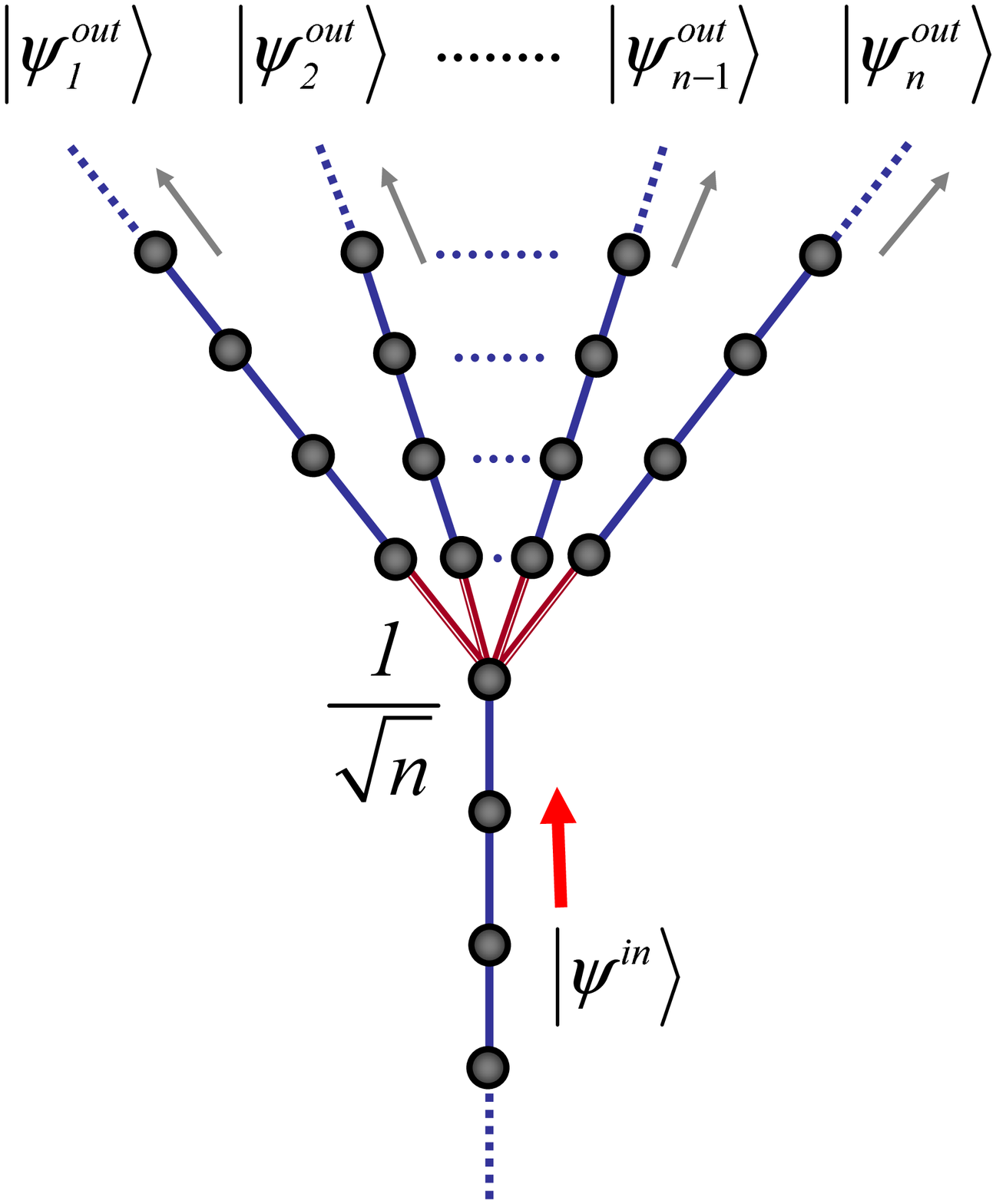} %
\includegraphics[ bb=37 143 554 758, width=4.0 cm, clip]{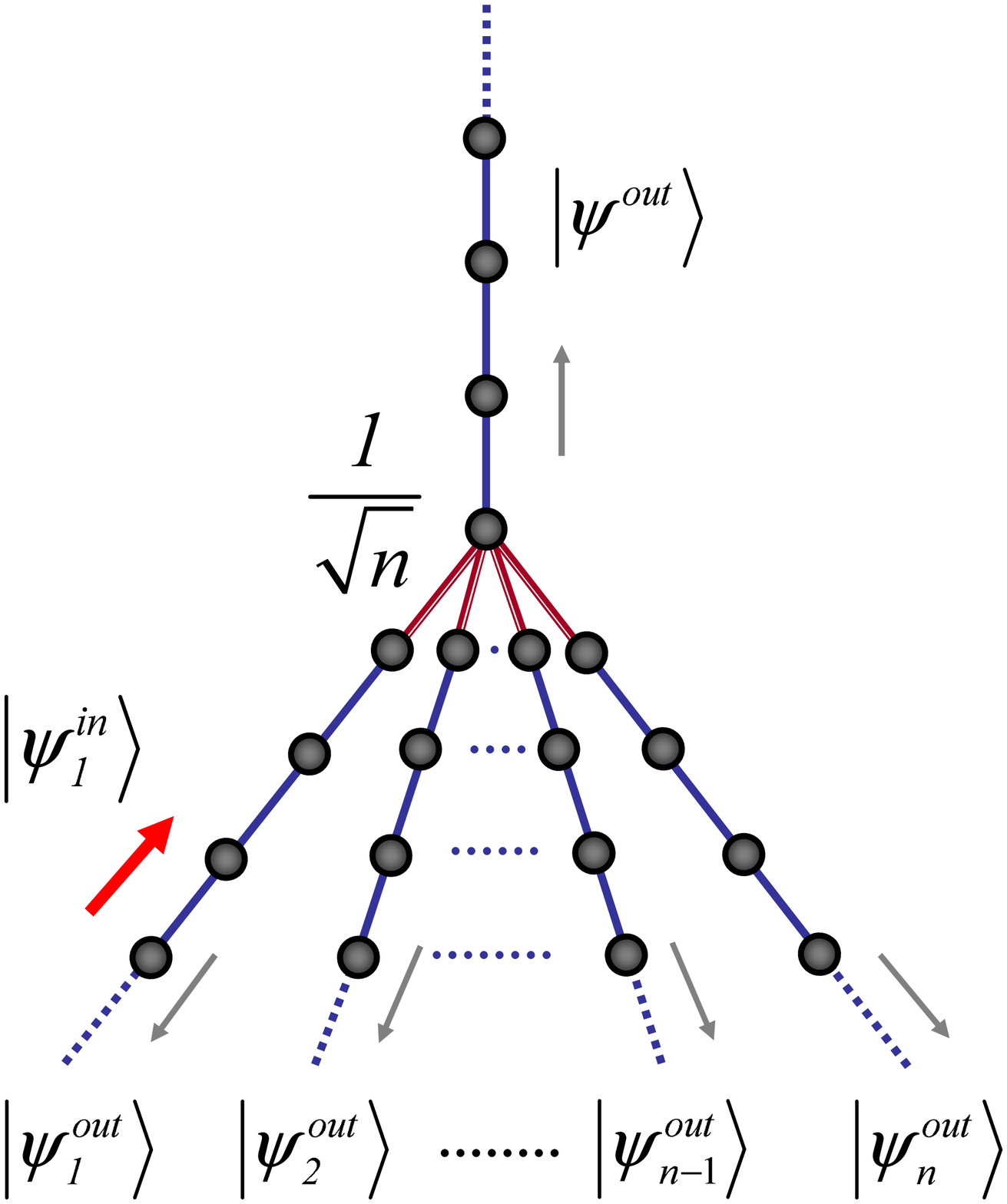}
\caption{\textit{(Color online) Schematic illustration for a }$1\times n$%
\textit{\ beam splitter. Left panel: an incident wave packet }$\vert \protect%
\psi ^{in}\rangle $\textit{, is split into }$n$\textit{\ sub-wave packets }$%
\vert \protect\psi _{j}^{out}\rangle $\textit{\ }$(j=1,...,n)$\textit{\
without any reflection. Right panel: an incident wave packet }$\vert \protect%
\psi _{l}^{in}\rangle $\textit{\ along one of the }$n$\textit{\ arms is
divided into }$n+1$\textit{\ sub-wave packets }$\vert \protect\psi %
_{j}^{out}\rangle $\textit{\ }$(j=1,...,n)$\textit{\ and }$\vert \protect%
\psi ^{out}\rangle $\textit{.}\emph{\ }}
\label{fig5}
\end{figure}

Similarly as shown in Ref. {\cite{Yang EPJB}}, we can perform the following
transformations%
\begin{eqnarray}
c_{i}^{\dag } &=&a_{i}^{\dag },\left( i\leq l_{a}\right) , \\
c_{i+l_{a}}^{\dag } &=&\frac{1}{\sqrt{n}}\sum_{j=1}^{n}b_{j,i}^{\dag
},\left( i\leq l_{b}\right) ,  \notag \\
d_{l,i}^{\dag } &=&\frac{1}{\sqrt{n}}\sum_{j=1}^{n}e^{-i\frac{2\pi l}{n}%
j}b_{j,i}^{\dag },\left( l=1,2,...,n-1\right) ,  \notag
\end{eqnarray}%
where $c_{i}^{\dag }$\ and $d_{l,j}^{\dag }$\ are also corresponding boson
or fermion operators. Under the transformations, Hamiltonian (\ref{Quantum
network}) can be written as

\begin{eqnarray}
\mathcal{H} &=&\mathcal{H}_{c}+\sum_{l=1}^{n-1}\mathcal{H}_{l}, \\
\mathcal{H}_{c} &=&-\frac{J^{\bot }}{2}\sum_{i=1}^{l_{a}+l_{b}-1}\left(
c_{i}^{\dag }c_{i+1}+\mathrm{H.c.}\right) ,  \notag \\
\mathcal{H}_{l} &=&-\frac{J^{\bot }}{2}\sum_{i=1}^{l_{b}-1}\left(
d_{l,i}^{\dag }d_{l,i+1}+\mathrm{H.c.}\right) ,  \notag \\
&&\left( l=1,2,..,n-1\right) .  \notag
\end{eqnarray}%
Note that the sub-Hamiltonians satisfy

\begin{equation}
\left[ \mathcal{H}_{c},\mathcal{H}_{l}\right] =0,\left[ \mathcal{H}_{l},%
\mathcal{H}_{m}\right] =0,
\end{equation}%
which indicate that the original quantum network, $1\times n$ beam splitter
can be\ decomposed into $n$ independent chains, one of them is the length of
$l_{a}+l_{b}$\ and the rest are all the length of $l_{b}$. Based on the fact
that a moving wave packet can propagate freely along the $n$ independent
chains, we have the following processes

\begin{eqnarray}
\left\vert \psi ^{in}\right\rangle &\longrightarrow &\frac{1}{\sqrt{n}}%
\sum_{j=1}^{n}\left\vert \psi _{j}^{out}\right\rangle , \\
\frac{1}{\sqrt{n}}\sum_{j=1}^{n}\left\vert \psi _{j}^{in}\right\rangle
&\longrightarrow &\left\vert \psi ^{out}\right\rangle ,  \notag \\
\sum_{j=1}^{n}e^{-i\frac{2\pi m}{n}j}\left\vert \psi _{j}^{in}\right\rangle
&\longrightarrow &\sum_{j=1}^{n}e^{-i\frac{2\pi m}{n}j}\left\vert \psi
_{j}^{out}\right\rangle ,  \notag \\
&&\left( m=1,2,..,n-1\right) .  \notag
\end{eqnarray}%
Through a straightforward algebra, we obtain a set of expressions
\begin{eqnarray}
\left\vert \psi _{l}^{in}\right\rangle &\longrightarrow &\frac{1}{\sqrt{n}}%
\left\vert \psi ^{out}\right\rangle -\frac{1}{n}\sum_{j=1}^{n}\left\vert
\psi _{j}^{out}\right\rangle +\left\vert \psi _{l}^{out}\right\rangle ,
\label{l_in} \\
&&\left( l=1,2,..,n\right) .  \notag
\end{eqnarray}%
Then for a given incident wave packet along any branch of a $1\times n$ beam
splitter, the probability amplitudes of all the output wave packets can be
obtained exactly.

\end{document}